# Data-Centric Safety and Ethical Measures for Data and AI Governance

Srija Chakraborty

Universities Space Research Association
schakraborty@usra.edu

## Abstract

Datasets play a key role in imparting advanced capabilities to artificial intelligence (AI) foundation models that can be adapted to various downstream tasks. These downstream applications can introduce both beneficial and harmful capabilities – resulting in dual use AI foundation models, with various technical and regulatory approaches to monitor and manage these risks. However, despite the crucial role of datasets, responsible dataset design and ensuring data-centric safety and ethical practices have received less attention. In this study, we propose a responsible dataset design framework that encompasses various stages in the AI and dataset lifecycle to enhance safety measures and reduce the risk of AI misuse due to low quality, unsafe and unethical data content. This framework is domain agnostic, suitable for adoption for various applications and can promote responsible practices in dataset creation, use, and sharing to facilitate red teaming, minimize risks, and increase trust in AI models.

## Introduction

Recent years have witnessed a tremendous increase in the capacity of artificial intelligence/ machine learning (AI/ML) methods. Specifically, when trained on large datasets using self-supervised learning and pre-training, the resulting foundation models learn meaningful representations of the datasets it was trained on and have demonstrated high generalization capabilities when adapted to a range of downstream tasks with both positive and negative impact (Bommasani et al. 2021, Bommasani et al. 2023). The improved performance has been attributed to the large size datasets, novel ML architectures in the transformer family, and large compute power used to train the models (Buchanan 2020), with various analyses of scaling laws extrapolating how these capabilities are expected to evolve. The role played by the size, variety, and quality of datasets have been critical at enabling foundation models to learn a generalized representation of different datasets, with a recent study highlighting that datasets used in training have been undisputable in foundation models adapting to both positive and nefarious use cases, resulting in dual-use capabilities (Gupta et al. 2024) but assessing data quality has not received enough attention (Sambasivan et al. 2021). Unsafe outcomes from AI use can be both intentional and unintentional and can introduce legal, ethical, security risks and the role of data in enabling this is noteworthy. These risks can be multifaceted including– copyright infringement (Longpre et al. 2024), fairness violation and discrimination (Bender et al. 2021), failing legal and compliance requirements, introduction of reputational risks by generating biased and/or toxic context and stereotypes (Navigli et al. 2023, Deshpande et al. 2023), misinformation, introduce cybersecurity-risks and jailbreaking capabilities, and assist in harmful tasks when used without guardrails with chemical and biological data tasks, and to develop lethal autonomous weapons (Urbina et al. 2022, Qi et al. 2023, Fu et al. 2024). Such outcomes can cause serious threats to public safety, erode democratic values, international humanitarian law, human rights, and reduce trust in AI. However, emphasis on disclosure for governance has focused on models and compute and less on data, even though data plays a key role. Recent policies introduce regulatory mandates on training data transparency (Irwin 2024) and without dataset release red teaming may be limited in understanding full range of model capabilities and anticipating risks. Assessing datasets and their content may also be essential for adding layers of data security, privacy protection measures in high-risk uses of AI (Qi et al. 2023, Fu et al. 2024 Li et al. 2024, Phuong et al. 2024).

Thus, defining standards, adopting best practices in dataset design, monitoring and evaluating dataset quality for suitability of training foundation models, fine-tuning to downstream applications and assessing how likely it is for a dataset to introduce risks, legal and ethical concerns into an AI model are necessary to facilitate red-teaming and for establishing trust in AI. Guidelines for sharing datasets already exist but require improvements to address different stages of the AI lifecycle for being robust to misuse in the current landscape with widespread use of AI. AI dataset transparency studies have recommended 'data cards' that record various details such as a dataset's upstream source, data collection and annotation, training and evaluation details, intended use shared as documentation (Gebru et al. 2021, Holland et al. 2022, Pushkarna et al. 2022, Akhtar et al. 2024). An open source, platform independent responsible AI metadata format, Croissant RAI (Jain et al. 2024) was introduced that produces machine-readable metadata to enhance responsible AI practices using these recommendations. Other studies examined transparency in foundation models' training datasets under the foundation model transparency index (Bommasani et al. 2023) that lists multiple data-related indicators that should be considered, while the NIST AI Risk Management Framework has recommended dataset transparency for risk management in three of its risk management axes, namely 'govern', 'risk mapping', and 'measure' axes (NIST 2024). While these studies have considerable overlap in identifying steps to improve dataset transparency(Gebru et al. 2021, Holland et al. 2022, Pushkarna et al. 2022, Jain et al. 2024), there are several additional transparency indicators specific to datasets for training large models.

In this study, we propose a multi-stage responsible dataset design framework that unifies existing studies and recommends best practices at each stage in the AI and dataset lifecycle to enhance dataset safety measures for preventing the misuse of AI due to low-quality, unsafe, and/or unethical content in datasets. At each stage in the lifecycle, the approach maps data-centric risks, outlines potential measures to monitor and manage these risks with the overarching goal of promoting AI governance. The approach is designed to maximize the usefulness of AI/ML (releasing datasets with safety measures that can be used in diverse applications) while minimizing harms and detecting risks early (through multi-stage safety measures). Our approach considers safety best practices at different stages in the dataset and AI/ML lifecycle and proposes a theoretical approach to facilitate red teaming and AI governance through managing datasets.

## Proposed Responsible Dataset Design Framework

Datasets used for training (pre-training or fine-tuning) AI models can go through multiple stages. These include (a) ideation and design, (b) data collection, (c) data pre-processing and AI-readiness, (d) training, validation, (e) finetuning, (f) release, and maintenance – with stages (d), (e), and (f) having direct overlap with AI model use. The proposed work examines dataset-specific safety measures at each of these stages. We note that these stages may not be linear and depending on the development pipeline, some stages may be redundant. For e.g., an AI company may use training data to train and release large pre-trained models, but not have a fine-tuning step (step e), or downstream application developers will require fine-tuning dataset but not require pre-training data (step d), while other development pipelines may include both pre-training and fine-tuning. Nonetheless, quality of data and its relevance for the applications of a trained model will be relevant for its impact on AI inference and outcomes. Thus, standardizing dataset development, collection, maintenance and disclosure remains crucial for transparent, safe and human-centric AI technologies. Moreover, datasets used in pre-training will impact the areas a model can be effectively adapted to and thus essential for downstream use case developers to understand even though they may not directly develop or use the pre-training dataset. Our proposed multi-stage measures for dataset safety and transparency are described below:

**(a) ideation and dataset design**: This step relates to the underlying motivation to create a dataset by various means such as surveys, sensor data collection, data scraping and mixing by building on previously generated content to support in training and developing an AI model. Responsible data design practices should be adopted at this stage that includes (i) defining the task to collect data for and its scope, (ii) ensuring the task is ethical, protects human rights, democratic values, (iii) identifying possible sources to collect data from, defining features to record, designing survey questionnaire, and (iv) examining existing legal and technical standards to uphold – for e.g. standards related to involving human subjects, guidelines on personally identifiable information (PII) use, reviewing and complying with data license and establishing agreements for new data collection, examining if export control regulations are necessary for sensitive data that may otherwise impact public safety, critical infrastructure and security.

**(b) data collection**: After identifying data collection sources and reviewing standards, this step involves data collection in a manner that upholds the identified standards. Responsible data collection practices for this stage of the pipeline involve (i) data collection strategy that does not violate any legal, technical, ethical standards and in case of high-risk datasets (related to security, public health and safety) the appropriate security requirements are to be followed, (ii) collecting dataset features and recording a quality indicator for quality assurance/quality checks (QA/QC) for each sample for any new data collection, (iii) for use of existing data, source data provenance and quality should be examined either by domain experts, or through any automated expert-guided data filtering methods, (iii) using informed consent practices to collect data, and disclose any data-opt out protocols at different stages in the AI lifecycle, and privacy preserving mechanisms to be used in the study, (iv) disclose ethical practices adopted in the study for human data annotators, (v) document all strategies decisions used in data collection including collection method, data source and copyrights, dataset features recorded, data and feature quality, licenses used while collecting data, ethical practices if human annotators were used to create labels.

Data quality recording is critical for ensuring that high quality inputs are used in training AI models and the quality records can be noted through data flags indicating reliability of inputs similar to science domains. Agreement in dataset labels between annotators if applicable and can be captured as metadata files to be shared as in commonly used formats such as json or yaml. Furthermore, it may be necessary to also ensure that datasets collected are representative for the task at hand and requires checking that the volume of high quality data needed to train a model for the different classes have been collected, and the steps can be iterated to fill in any gaps. This step requires socio-technical expertise and multidisciplinary inputs should be gathered to mitigate any biases. Datasets should be stored in a secure manner with access control (blocklists and allowlists), especially for sensitive and high-risk content.

**(c) dataset pre-processing and AI-readiness:** At this stage the collected data is processed for use with AI models. Responsible and safety-aware measures for dataset design include (i) checking that any potential bias, harmful or toxic content, stereotypes are filtered out using existing ML approaches to detect bias, toxic context, (ii) checking if dataset contains high-risk content (chemical, biological sequences and their properties) that can jeopardize public safety and be misused for dual-use purposes and are to be filtered or only shared with individuals with clearance to handle such information for research (iii) use of data quality measures, label agreements recorded in step (b) to filter out low-confidence inputs whenever possible, (iv) exploring dataset balance and class representation through exploratory data analysis, dataset clustering to ensure that all classes are appropriately represented and iterating step (b) if necessary. For carrying out these tasks, this step should also include multidisciplinary oversight and analytical tools for examining steps (i) to (iv) and can be shared as APIs with red teaming and AI assurance experts. Documentation of all steps, results and evaluation, and class representatives, distributions of recommended training, validation, test splits should also be recorded at this stage. Synthetically generated data samples are also used in training AI-foundation models and can adversely affect model performance (Shumailov et al. 2023). Its use should also be indicated in the documentation along with flagging synthetic samples in the dataset.

**(d and e) training evaluation, and fine-tuning:** In this stage, the processed, AI-ready dataset is used to train different AI model architectures based on the task at hand. While responsible and safety-aware measures primarily include monitoring model performance and its propensity to cause harm, various data-centric measures can also be used to assess risk. As these measures apply to both model training and fine-tuning stages, we discuss these steps jointly. Specifically, these measures include (i) using guidelines on training, validation, test splits, data quality, synthetic data indicated from step (c) to train models along with noting possible harmful, unsafe content that the dataset has been checked and filtered for, (ii) assessing model performance on various unethical and unsafe prompts based on training data content using publicly available benchmarks (Li et al. 2024) or expert curated prompts – including cyberattack prompts, chemical, biological sequence generation tasks, optimization tasks to assist with planning illegal operations, coercion, disinforming users, eroding trust in democratic values and institutions, (iii) assessing model performance to preserve and guarantee fairness in various applications and can be necessary when AI methods are embedded in public interest technologies and data consists protected categories, (iv) steps (ii) and (iii) are red teaming steps and as such require large multidisciplinary expertise (e.g. AI/ML, fairness, cyber-risk, security and autonomous systems, chemical-biological systems, social sciences, ethics, etc.) and require careful evaluation using measures that reflect both technical performance and social and safety metrics, (v) detected threats from steps (ii) and (iii) should be addressed by embedding safety alignment methods, filtering unacceptable outputs, performing AI model and data ablation experiments to detect underlying source of harmful content – for e.g. specific data samples, features, low quality data, use of explainable AI/ML methods to examine features, samples contributing to negative outcomes should be explored and post-decision filtering should be applied.

The safety-aware steps for training and fine-tuning share a high overlap. While the training step (d) should be evaluated for a broad range of high-risk and unethical uses, similar measures are also applicable for fine-tuning step (e) but on a narrow, targeted set of tasks with lower resources needed for evaluation. For red-teaming and assessing model and data concurrently, these components should undergo phased release with access given initially only to experts, researchers and upon meeting acceptable performance prepared for release. All steps, evaluation measures, performance benchmarks, decisions should be documented. Furthermore, AI model developers should clearly define what the models and datasets used to train it are appropriate for by listing possible use cases and describe safe and ethical applications of the model and datasets. Based on the assessment from this step, both models and datasets should be released with appropriate license and documentation whenever possible, and with access controls (access list – list of safe users, blocklist – list of unsafe users) whenever necessary, especially when datasets contain sensitive information or there is a scope of misuse. While it may be possible to block release of high-risk models and datasets altogether, ensuring countermeasures to reduce risk can allow maximizing adaptation for positive uses of AI by adding appropriate security, access control checks, monitoring misuse, and improving preparedness to respond to misuse.

**(f) release and maintenance:** Once a model and dataset has been assessed for risk and can be released, additional measures are essential at this step to minimize misuse. The measures for dataset safety can differ from those for model safety and we discuss the measures for enhancing dataset safety once released. Release of datasets are essential as it promotes transparency for evaluating AI models, supports AI governance, facilitates further research and analysis reusing the dataset to explore other beneficial use cases, and may be mandatory to disclose based on newly introduced policies (Irwin 2024). The specific steps for responsible use of AI datasets post-release are (i) sharing dataset appropriately based on content (from analysis of steps (a),

(b), (c)) and tendency of the dataset to introduce unacceptable capabilities to AI models for generating harmful, unsafe, unethical outputs (from analysis of steps (d) and (e)) – these could include decisions on licensing, export control, user list monitoring (through blocklists, allowlists), privacy preserving measures, etc., (ii) thorough documentation of data through all lifecycle stages, including decisions made, sources, quality, labels, class distribution, statistics from step (c) (i) - (iv) summarizing dataset properties, capabilities arising from it observed through trained models in steps (d) and (e), cleaning, filtering, alignment measures, performance with data ablation methods, and training, validation, test splits used along with metrics determined at red teaming (although these steps perform a coupled data and model assessment), (iii) descriptions of why and how the data was collected, ethical, legal and technical compliance that was adhered to, (iv) descriptions of acceptable use cases and liability statements in the event of violating this list to train other models with the released dataset and intentionally or unintentionally introducing unacceptable outcomes in high-risk or high consequence uses, (v) tracking data downloads, use and updating block- and allow-lists, and (vi) sharing user guides and APIs to download, access, explore data when possible and statements on updates to data, data versioning.

The datasets developed should be shared in platform-independent, interoperable file formats along with the described documentation, metadata, user guides and APIs to promote beneficial outcomes using AI models, accelerate research while minimizing risk and safety concerns. A high level summary or checklist of datasets outlining type of data, its volume, upstream licenses of source data, high risk (chemical biological sequences, cyber risk, autonomous weapons information, geopolitical disinformation), other sensitive information (PII data, human-centric information), fairness, bias, QA/QC checks undergone by the data, red-teaming and safety performance, and license for use and access should also be disclosed by the creators by assembling information from all stages in the framework.

Although responsible AI data formats are being developed such as Croissant Responsible AI (Jain et al. 2024), Data Nutrition Labels (Holland et al. 2021), Dataset cards (Gebru et al. 2021), in a rapidly evolving field with new capabilities, multi-faceted measures are necessary for ensuring safety from datasets such as those using foundation models and generative AI. These considerations are yet to be included in these formats. The capabilities outlined in this study can be embedded into these responsible AI-relevant data formats and documentation to reuse existing structures and rapidly develop dataset standards for promoting safety and assisting with AI governance and regulatory checks. Additionally, the approach would also require stacking any domain-specific (such as privacy standards in medical domain) and geography-specific (US, EU, African Union, Latin American, and other national, state regulation) safety requirements to ensure compliance and minimize chances of introducing risks.

**Limitations:** Although this study explores a multi-stage approach throughout the data and AI lifecycle to introduce safety-enhancing measures, there are AI safety concerns that are not addressed through this approach. Firstly, this approach does not address risks introduced through open model weight sharing that can be fine-tuned for nefarious purposes. Secondly, documenting all steps at each stage and all associated decision making (including licenses, ethical legal standards, data collection methods, features selected, bias in annotating data, prompt generation and red teaming) requires significant effort and expertise and it may be challenging to team up these experts. Thirdly, regulatory and compliance measures vary across disciplines and regions that can be challenging to meet through a single framework and may require a tiered approach by building on the proposed framework. Finally, adopting the proposed approach may require additional effort. Such concerns can be addressed by using existing documentation styles and metadata formats (Akhtar et al. 2024) or other common tools in different domains and reducing effort needed in building responsible datasets.

## Conclusion

The proposed study introduces a multi-stage responsible dataset design framework to ensure safety measures for developing, maintaining and sharing datasets used in training AI models. The framework is designed with the objective of enhancing transparency, trust, promoting beneficial uses of AI and associated datasets while minimizing risks from intentional and unintentional misuse. While AI models have shown rapid improvement in performance and have been scrutinized, the role of datasets have received less attention. The proposed approach examines the data and AI lifecycle, maps risks at each of these stages introduced through misuse of data, and outlines measuring, documentation strategies to manage these risks. The recommendations can be introduced into existing responsible AI dataset efforts to reuse existing structures and promote rapid progress towards standardizing AI dataset attributes to increase safety. The framework is generic and can be adapted to various disciplines to promote responsible dataset development and maintenance and support data and AI governance.

## References

Akhtar, M.; Benjelloun, O.; Conforti, C.; Giner-Miguelez, J.; Jain, N.; Kuchnik, M.; Lhoest, Q.;
Marcenac, P.; Maskey, M.; Mattson, P.; Oala, L.;Ruyssen, P.; Shinde, R.; Simperl, E.; Thomas, G.;Tykhonov, S.; Vanschoren, J.; Vogler, S.; and Wu, C.-J.2024. Croissant: A Metadata Format for ML-Ready Datasets. In Proceedings of the Eighth Workshop on Data Management for End-to-End Machine Learning, DEEM '24. New York, NY, USA: Association for Computing Machinery. ISBN 979840070611.

Bender, E.M., Gebru, T., McMillan-Major, A. and Shmitchell, S., 2021, March. On the dangers of stochastic parrots: Can language models be too big?🦜. In *Proceedings of the 2021 ACM conference on fairness, accountability, and transparency* (pp. 610-623).

Bommasani, R., Hudson, D.A., Adeli, E., Altman, R., Arora, S., von Arx, S., Bernstein, M.S., Bohg, J., Bosselut, A., Brunskill, E. and Brynjolfsson, E., 2021. On the opportunities and risks of foundation models. *arXiv preprint arXiv:2108.07258*.

Bommasani, R., Klyman, K., Longpre, S., Kapoor, S., Maslej, N., Xiong, B., Zhang, D. and Liang, P., 2023. The foundation model transparency index. *arXiv preprint arXiv:2310.12941*.

Buchanan, B., 2020. The AI triad and what it means for national security strategy. *Center for Security and Emerging Technology*.

Deshpande, A., Murahari, V., Rajpurohit, T., Kalyan, A. and Narasimhan, K., 2023. Toxicity in chatgpt: Analyzing persona-assigned language models. arXiv preprint arXiv:2304.05335.

Fu, X., Li, S., Wang, Z., Liu, Y., Gupta, R.K., Berg-Kirkpatrick, T. and Fernandes, E., 2024. Imprompter: Tricking LLM Agents into Improper Tool Use. *arXiv preprint arXiv:2410.14923*.

Gebru, T.; Morgenstern, J.; Vecchione, B.; Vaughan, J. W.; Wallach, H.; Iii, H. D.; and Crawford, K. 2021. Datasheets for datasets. Communications of the ACM, 64(12): 86–92.

Gupta R, Walker L, Corona R, Fu S, Petryk S, Napolitano J, Darrell T, Reddie AW. Data-Centric AI Governance: Addressing the Limitations of Model-Focused Policies. arXiv preprint arXiv:2409.17216. 2024 Sep 25.

Holland, S.; Hosny, A.; Newman, S.; Joseph, J.; and Chmielinski, K. 2018. The Dataset Nutrition Label: A Framework To Drive Higher Data Quality Standards.arXiv:1805.03677.

Irwin J, "AB 2013, Generative artificial intelligence: training data transparency.", 2024, https://leginfo.legislature.ca.gov/faces/billNavClient.xhtml?bill_id=202320240AB2013, [online accessed: Nov 24, 2024]

Jain, N., Akhtar, M., Giner-Miguelez, J., Shinde, R., Vanschoren, J., Vogler, S., Goswami, S., Rao, Y., Santos, T., Oala, L. and Karamousadakis, M., 2024. A Standardized Machine-readable Dataset Documentation Format for Responsible AI. *arXiv preprint arXiv:2407.16883*.

Li, N., Pan, A., Gopal, A., Yue, S., Berrios, D., Gatti, A., Li, J.D., Dombrowski, A.K., Goel, S., Phan, L. and Mukobi, G., 2024. The wmdp benchmark: Measuring and reducing malicious use with unlearning. *arXiv preprint arXiv:2403.03218*.

Longpre, S., Mahari, R., Lee, A.N., Lund, C.S., Oderinwale, H., Brannon, W., Saxena, N., Obeng-Marnu, N., South, T., Hunter, C.J. and Klyman, K., 2024, January. Consent in crisis: the rapid decline of the AI data commons. In *The Thirty-eight Conference on Neural Information Processing Systems Datasets and Benchmarks Track*.

Navigli, R., Conia, S. and Ross, B., 2023. Biases in large language models: origins, inventory, and discussion. ACM Journal of Data and Information Quality, 15(2), pp.1-21.

NIST AI RMF, https://airc.nist.gov/docs/NIST.AI.600-1.GenAI-Profile.ipd.pdf , 2024

Phuong, M., Aitchison, M., Catt, E., Cogan, S., Kaskasoli, A., Krakovna, V., Lindner, D., Rahtz, M., Assael, Y., Hodkinson, S. and Howard, H., 2024. Evaluating frontier models for dangerous capabilities. *arXiv preprint arXiv:2403.13793*.

Pushkarna, M.; Zaldivar, A.; and Kjartansson, O. 2022. Data cards: Purposeful and transparent dataset documentation for responsible ai. In Proceedings of the 2022 ACM Conference on Fairness, Accountability, and Transparency, 1776–1826.

Qi, X., Zeng, Y., Xie, T., Chen, P.Y., Jia, R., Mittal, P. and Henderson, P., 2023. Fine-tuning aligned language models compromises safety, even when users do not intend to!. *arXiv preprint arXiv:2310.03693*.

Shumailov, I., Shumaylov, Z., Zhao, Y., Gal, Y., Papernot, N. and Anderson, R., 2023. The curse of recursion: Training on generated data makes models forget. arXiv preprint arXiv:2305.17493.

Urbina, F., Lentzos, F., Invernizzi, C. and Ekins, S., 2022. Dual use of artificial-intelligence-powered drug discovery. *Nature machine intelligence*, *4*(3), pp.189-191.